\documentclass{appolb}
\usepackage{graphicx}
\usepackage{amsmath}

\begin{document}
\title{MMHT PDFs: updates and outlook
\thanks{Presented at XXIII Cracow Epiphany Conference. Speaker: L. A. Harland-Lang.}%
}
\author{L. A. Harland-Lang, R. Nathvani, R. S. Thorne
\address{Department of Physics and Astronomy, University College London, WC1E 6BT, UK}
\\
{A. D. Martin
}
\address{Institute for Particle Physics
  Phenomenology, Durham University, Durham DH1 3LE, UK}
}
\maketitle
\begin{abstract}
\noindent We present the latest results of studies within the MMHT PDF framework. We discuss the impact of the most recent ATLAS 7 TeV jet data, demonstrating that while a good fit can be achieved for individual jet rapidity bins, it is not possible to achieve a good description of the data when all bins are fitted. We examine the role that the experimental correlated systematic uncertainties play in this, and demonstrate that by simply decorrelating no more than two sources of error between rapidity bins a remarkably improved description of the data can be achieved. We then study the impact of NNLO corrections, showing that a mild decrease in the fit quality is produced. We also present the results of including new LHC $W$, $Z$, $W+c$ and $t\overline{t}$ data on the MMHT14 PDF set, showing that a marked decrease in the $s+\overline{s}$ uncertainty is in particular achieved. Finally, some discussion of the latest work towards the inclusion of the photon PDF within the MMHT framework is presented.

\end{abstract}
  
\section{Introduction}

The MMHT14 parton distribution function (PDF) set~\cite{Harland-Lang:2014zoa} is the successor to the MSTW08~\cite{Martin:2009iq} PDFs. It combines a range of theoretical updates with new data, including for the first time measurements from the LHC. Subsequent studies on the $\alpha_s$ determination~\cite{Harland-Lang:2015nxa} and heavy quark mass dependence~\cite{Harland-Lang:2015qea} were performed. More recently, the impact of the final combined HERA I+II data set~\cite{Abramowicz:2015mha} was examined~\cite{Harland-Lang:2016yfn}. Here, the MMHT14 PDFs were found to give a good description of the HERA data, with the central values and uncertainties being changed relatively little by their inclusion in the fit. It was therefore decided not to release a new set at this point, but rather to wait until theoretical developments such as the full NNLO calculation of the jet production cross section were complete, and a more precise and varied range of LHC data became available. Recently there has also been great progress in the determination of the photon PDF~\cite{Martin:2014nqa,Harland-Lang:2016kog,Manohar:2016nzj}, with in particular the study of~\cite{Manohar:2016nzj} demonstrating that this object can be determined with percent--level precision. However, these findings have yet to be included in the context of a global PDF fit.

In these proceedings we report on work in all of the directions described above. Namely, we discuss a first look at the impact of LHC jets at NNLO, as well new LHC data on $W$, $Z$, $W+c$ and $t\overline{t}$ production on the PDFs, and present the latest work towards including a precisely determined photon PDF within the MMHT framework.

\section{The impact of LHC jet data}

Jet collider data play an important role in constraining the gluon PDF at higher $x$ and indeed in the past have placed the only reasonable direct constraint in this region (LHC measurements such as $t\overline{t}$ production, $Z$ and $W$ boson $p_\perp$ distributions and isolated photon production will also play an important role in the future). 
However a full NNLO calculation of jet production has until recently not been available. For this reason, in the MMHT14 set Tevatron jet data were included in the NNLO fit including the approximate threshold corrections of~\cite{Kidonakis:2000gi}, with the argument made that the difference between this and the full NNLO result should be under control, and in particular smaller than the experimental systematic uncertainties. Such a conclusion does not however follow in general at the LHC, where the larger $\sqrt{s}$ implies that much of the data lie very far from threshold, while those data that do not in fact probe a kinematically very similar region to the existing Tevatron data. For this reason LHC jet data were omitted from the fit, although a number of exploratory studies with different toy models for the NNLO K--factors were performed in~\cite{Harland-Lang:2014zoa}, and the impact on the gluon PDF was found to be relatively minor. 

However, in~\cite{Currie:2016bfm} the first calculation of fully differential jet production at NNLO was presented, allowing LHC jet data to be correctly included in a NNLO PDF fit for the first time. In this study NNLO K--factors are presented corresponding to the ATLAS 7 TeV measurement~\cite{Aad:2014vwa}, with jet radius $R=0.4$, and therefore in these proceedings we consider only this data set.

\subsection{NLO comparison and decorrelation study}

\begin{figure}[htb]
\centerline{%
\includegraphics[scale=0.5]{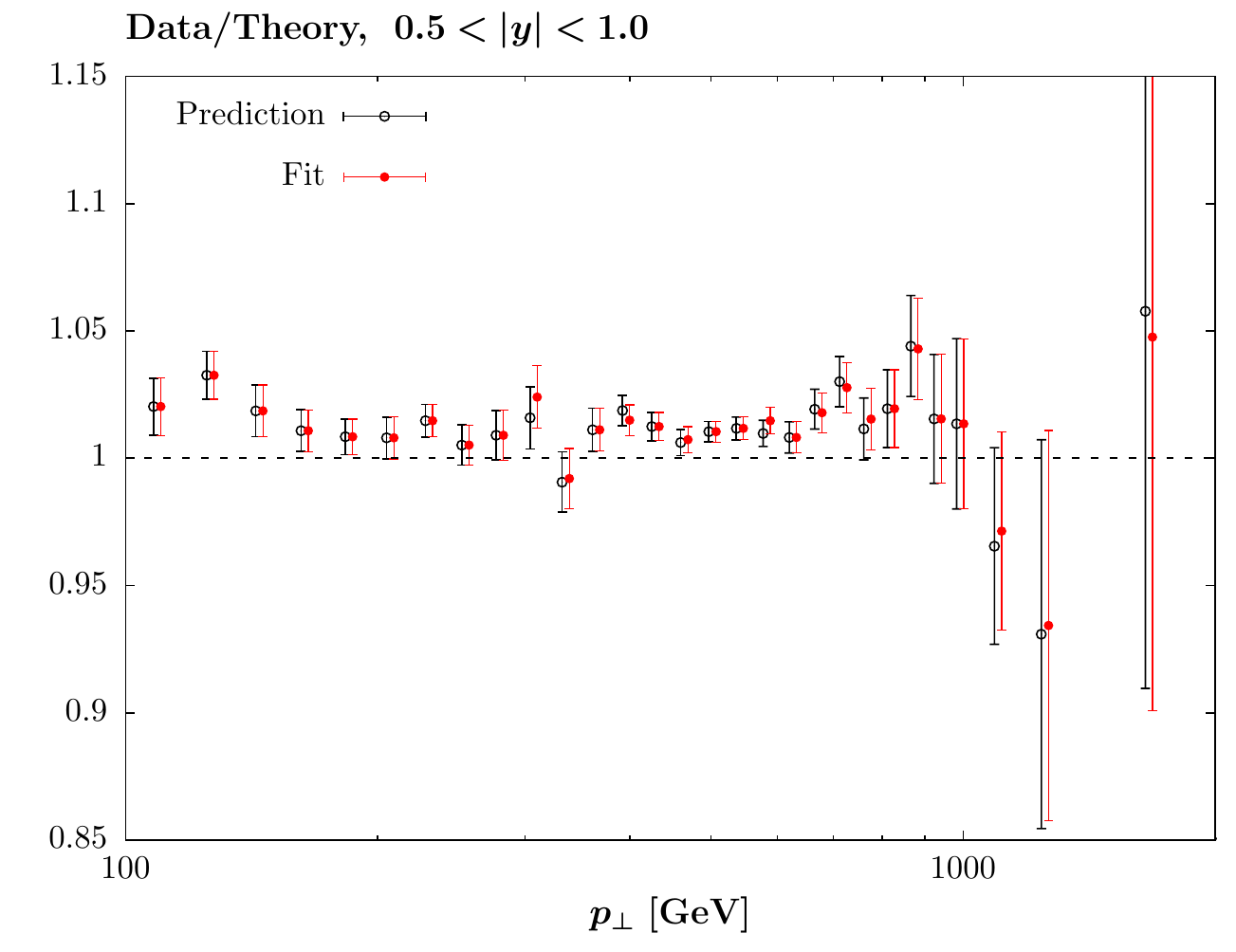}
\includegraphics[scale=0.5]{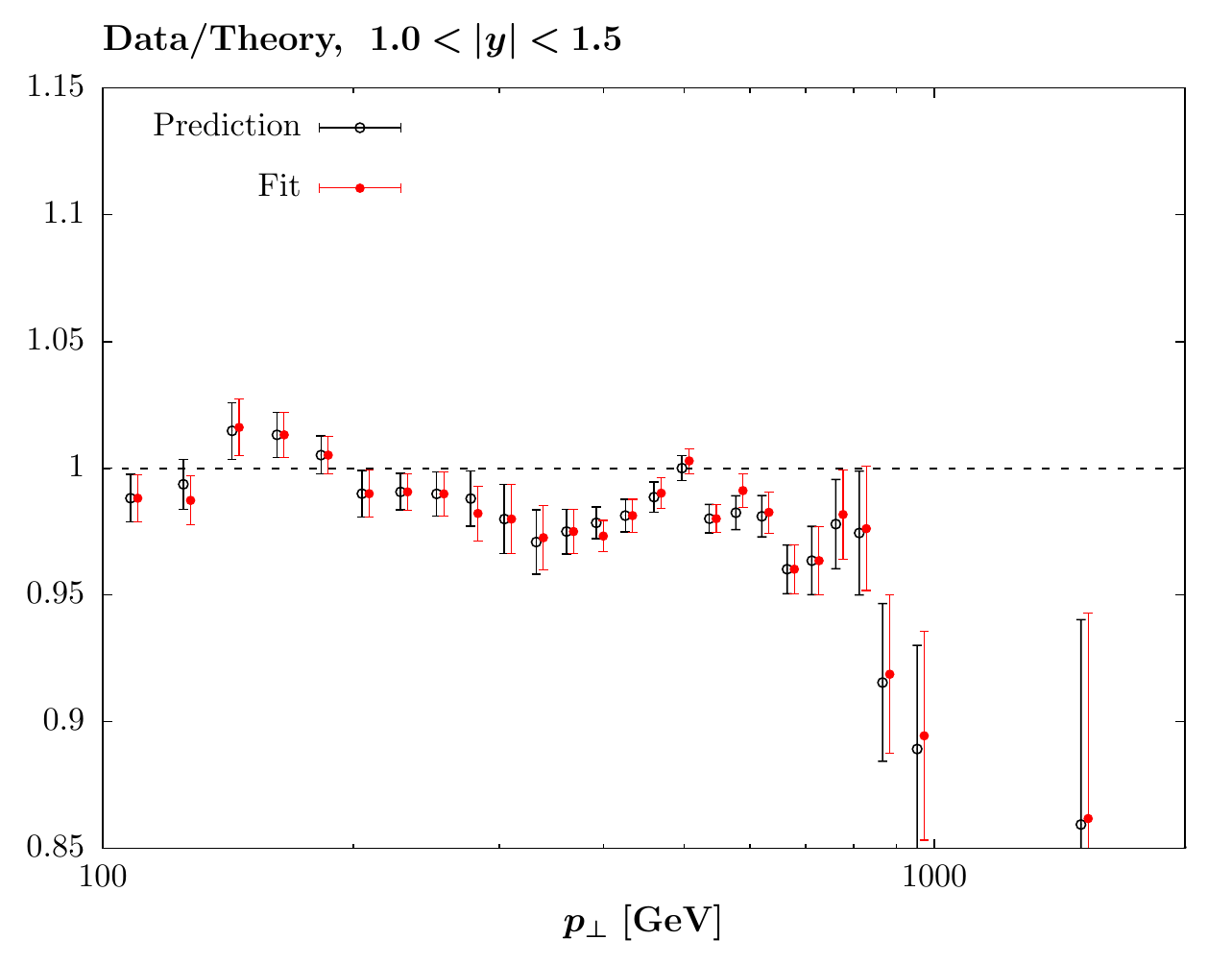}}
\caption{Comparison of NLO prediction and fit to ATLAS jet data~\cite{Aad:2014vwa} for two jet rapidity bins. Data/theory is plotted, with the data already shifted by the systematic uncertainties in order to achieve the best description. The displayed errors are purely statistical.}
\label{fig:jet1}
\end{figure}

We begin by considering the prediction and fit at NLO, before including the NNLO corrections. As the baseline PDF we use the MMHT14 set including the HERA I+II combined data~\cite{Abramowicz:2015mha}, that is as presented in~\cite{Harland-Lang:2016yfn}. The predicted and fit data/theory for the $0.5<|y_j|<1.0$ and $1.0<|y_j|<1.5$ jet rapidity bins are shown in Fig.~\ref{fig:jet1}, with the shifts due to the correlated systematic uncertainties included. The description of the data is visibly poor, and does not improve greatly with refitting. In particular, the $\chi^2$ for the description is $413$, decreasing to $400$ after refitting, for $140$ data points. From Fig.~\ref{fig:jet1} we can see that a significant contributing factor to this is an essentially systematic offset in the data/theory between the two neighbouring rapidity bins, but in opposite directions. As these probe PDF sets of the same flavour in very similar $x$ and $Q^2$ regions little improvement is possible (or observed) by refitting to this data. 

\begin{figure}[htb]
\centerline{%
\includegraphics[scale=0.4]{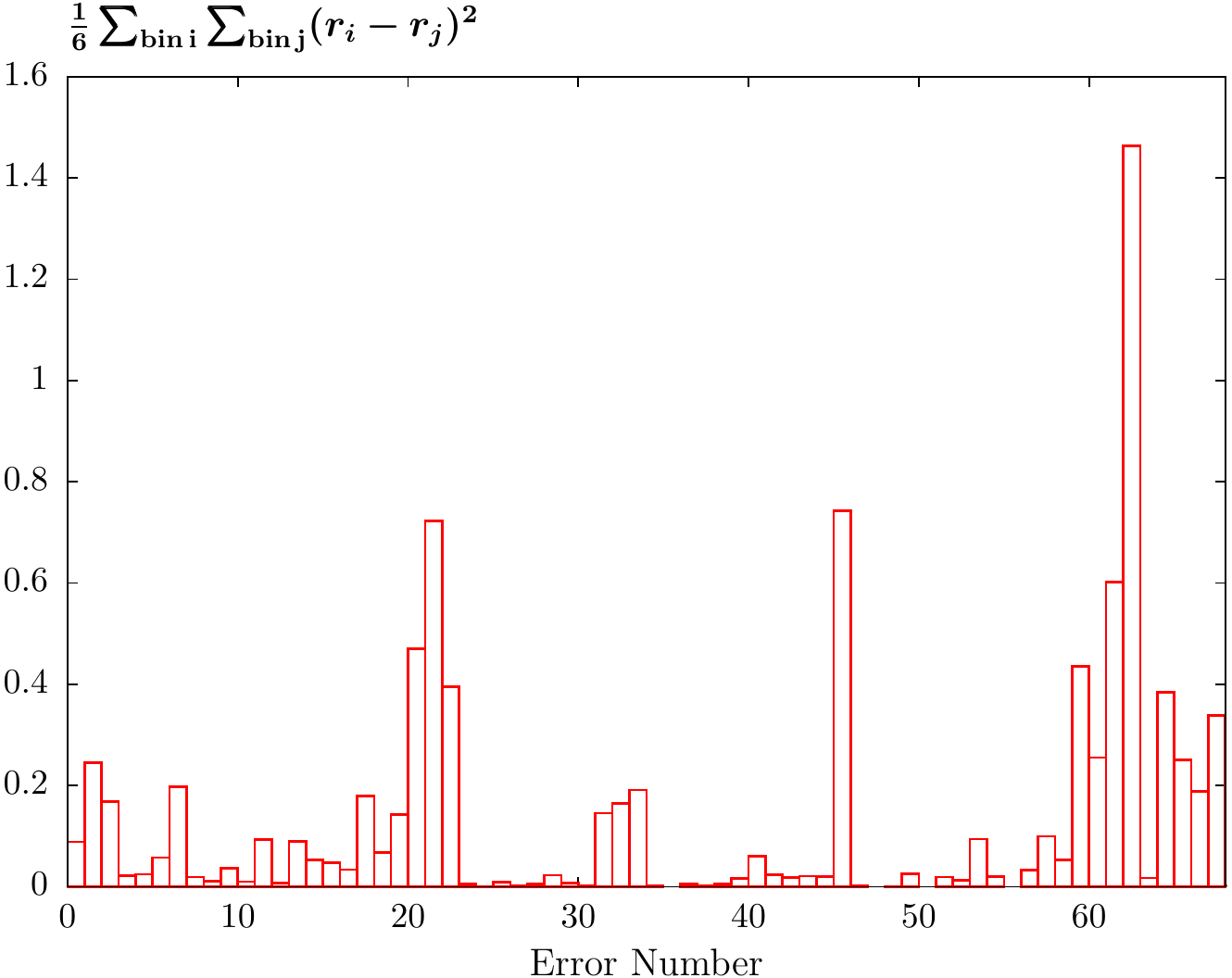}}
\caption{Average squared sum of the systematic shift differences $(r_i-r_j)^2$ for the first four rapidity bins of the ATLAS 7 TeV jet data~\cite{Aad:2014vwa}.}
\label{fig:jetshifts}
\end{figure}

The cause of this appears to lie with the shift allowed by the correlated systematic uncertainties. The ATLAS data contain a large number of individual correlated errors which are generally completely dominant over the (small) statistical errors; for the `weaker' assumption about error correlations defined in~\cite{Aad:2014vwa} that we take, there are 71 individual sources of systematic error. If we simply assume that all of these uncertainties are completely decorrelated between the six rapidity bins (while remaining fully correlated within the bins) a universally good description is found: in this case, the extra freedom allows the data to shift in order to achieve a reasonable data/theory description. This is however clearly a hugely over--conservative assumption. To be more precise, we examine the size of the shifts $r_k$ for each source of systematic uncertainty by which the theory (or equivalently, data) points are allowed to move, as defined in the $\chi^2$
\begin{equation}
\chi^2=\sum_{i=1}^{N_{\rm pts}}\left(\frac{D_i+\sum_{k=1}^{N_{\rm corr}}r_k\sigma_{k,i}^{\rm corr}-T_i}{\sigma_i^{\rm uncorr}}\right)+\sum_{k=1}^{N_{\rm corr}} r_k^2\;,
\end{equation}
where $D_i$ is $i$th data point, $T_i$ is the theory prediction and $\sigma^{\rm uncorr}_i$ ($\sigma^{\rm corr}_{k,i}$) are the uncorrelated (correlated) errors.
We in particular evaluate the shifts for each of the first four rapidity bins (from 0 to 2.0 in steps of 0.5) individually; including the last two rapidity bins, where the data tend to be less precise, does not affect the conclusions that follow. Any tensions between the different bins may then show up through significantly different $r_k$ values being preferred in the different rapidity bins, in order to achieve good individual fits. In Fig.~\ref{fig:jetshifts} we show the average squared sum of the shift differences $(r_i-r_j)^2$ for the four bins. It is clear that for a small subset  of the shifts the size of this difference is significantly larger than zero, indicating a large degree of tension. 

\begin{figure}[htb]
\centerline{%
\includegraphics[scale=0.4]{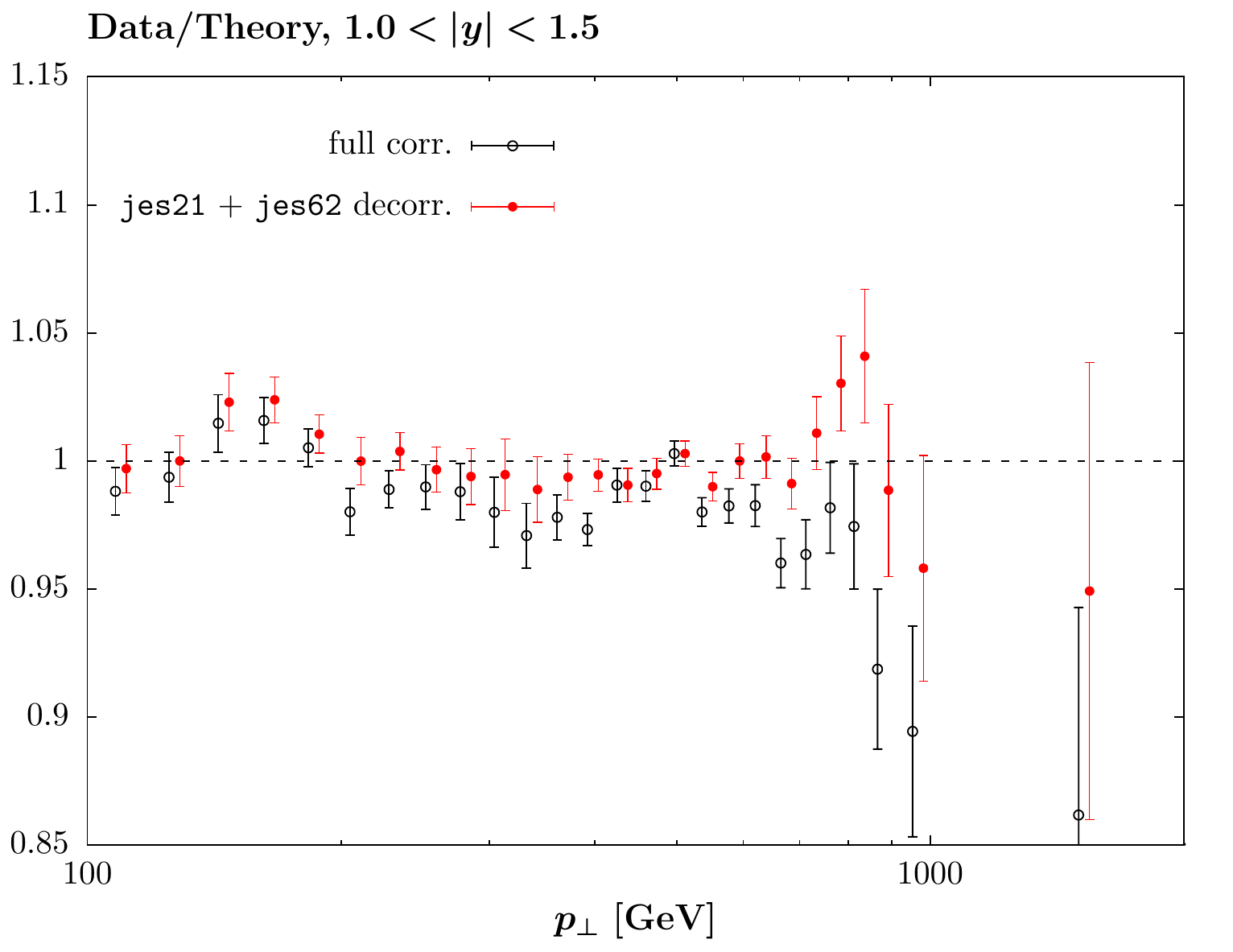}
\includegraphics[scale=0.4]{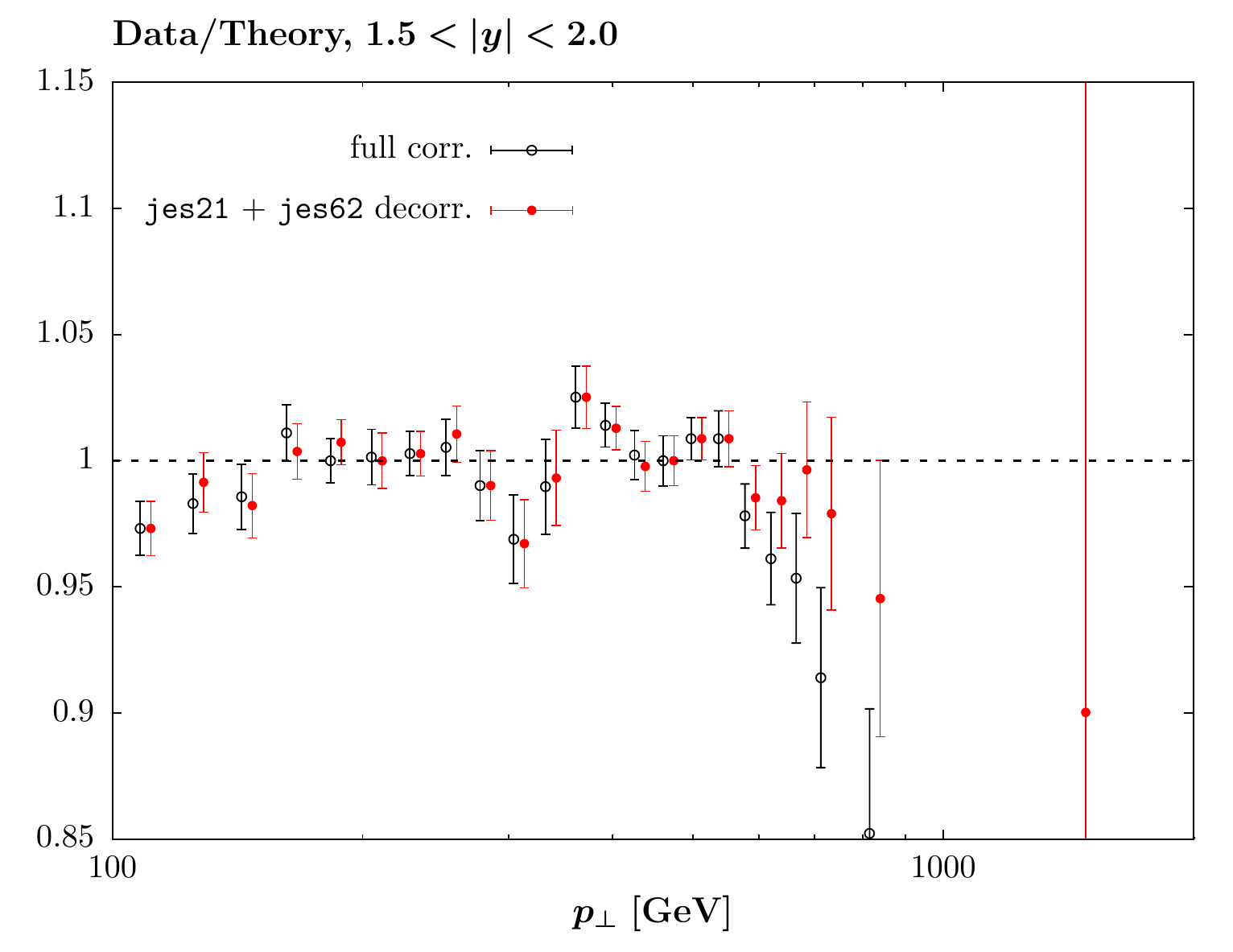}}
\caption{Data/theory fit as in Fig.~\ref{fig:jet1}, with and without the labelled systematic errors decorrelated between jet rapidity bins.}
\label{fig:jet2}
\end{figure}

The three shifts \texttt{jes21}, \texttt{45} and \texttt{62} as defined in~\cite{Aad:2014bia}, which correspond~\cite{ATLASpriv} to the multi-jet balance asymmetry, an in--situ statistical uncertainty and the jet energy scale close by jets, respectively, show particularly large differences. We therefore investigate the impact of decorrelating these systematic uncertainties alone between rapidity bins. The result for a selection of these three systematic uncertainties, as well as combinations of them, is shown in Table~\ref{tab:decor},  and is found to be dramatic. Simply decorrelating \texttt{jes21}, for example, leads to a reduction of $180$ points in $\chi^2$, giving almost a factor of 2 decrease in the $\chi^2/N_{\rm pts.}$ from 2.85 to 1.58; the result for the other two uncertainties is also significant, although not as large. Decorrelating \texttt{jes62} in addition gives a $\chi^2/N_{\rm pts.}$ of 1.27. The same data/theory comparisons as in Fig.~\ref{fig:jet1}, but including this decorrelation of \texttt{jes21} and \texttt{jes62}, are shown in Fig.~\ref{fig:jet2} and are visibly improved, with the additional freedom allowing the data/theory to shift in the different rapidity bins and achieve a good overall description. The correlation between systematic errors should clearly be determined by physics considerations and not simply the possibility of improving the theory description of the data\footnote{For an in--situ statistical uncertainty such as \texttt{jes45}, the correlations are particularly well determined~\cite{ATLASpriv}.}. Nonetheless this is an interesting finding which may hopefully guide a future re--analysis of the ATLAS systematic uncertainties, given the apparent tensions that are  present. For current purposes it will be useful to use this choice of error decorrelation when considering the impact of NNLO corrections in the following section.

\begin{table}[h]
\begin{center}
\begin{tabular}{|c|c|c|c|c|c|c|}
\hline
&Full&\texttt{21}&\texttt{45}&\texttt{62}&\texttt{21,62}&\texttt{21,45,62}\\
\hline
$\chi^2/N_{\rm pts.}$&2.85&1.58&2.26&2.36&1.27&1.23\\
\hline
\end{tabular}
\caption{$\chi^2$ per number of data points for fit to ATLAS jets data~\cite{Aad:2014vwa}, with the default systematic error treatment (`full') and with certain errors, defined in the text, decorrelated between jet rapidity bins.}\label{tab:decor}
\end{center}
\end{table}

\subsection{NNLO comparison}

\begin{table}[h]
\begin{center}
\begin{tabular}{|c|c|c|}
\hline
&Full corr.&\texttt{jes21,62} decorr.\\
\hline
$\chi^2$, NLO&(413) 400& (180) 178\\
\hline
$\chi^2$, NNLO & (443) 427 & (211) 204\\
\hline
\end{tabular}
\caption{$\chi^2$  for (description)fit to ATLAS jets data~\cite{Aad:2014vwa}, with the default systematic error treatment and with errors \texttt{jes21,62}, defined in the text, decorellated between jet rapidity bins.}\label{tab:nnlo}
\end{center}
\end{table}

\begin{figure}[htb]
\centerline{%
\includegraphics[scale=0.4]{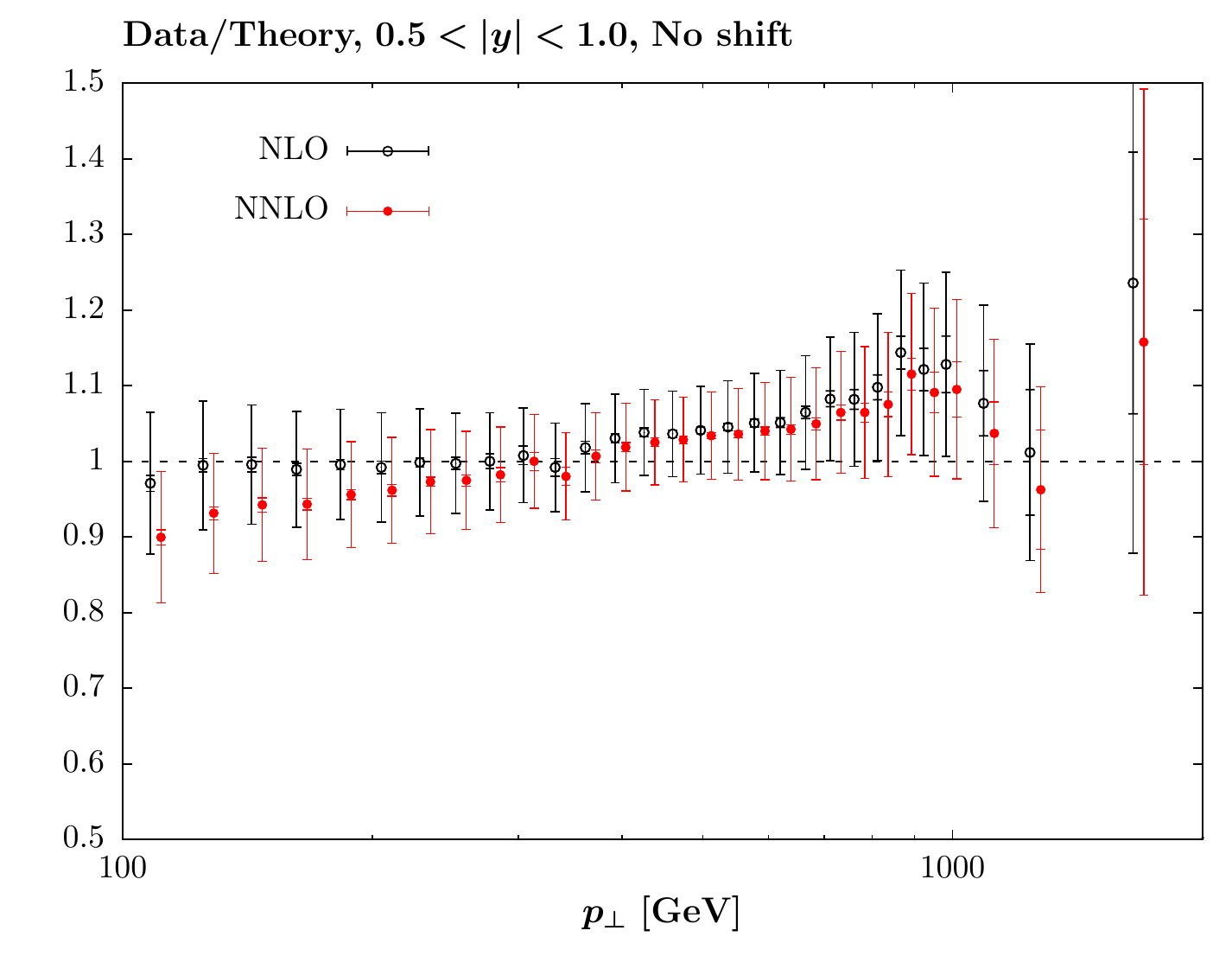}
\includegraphics[scale=0.4]{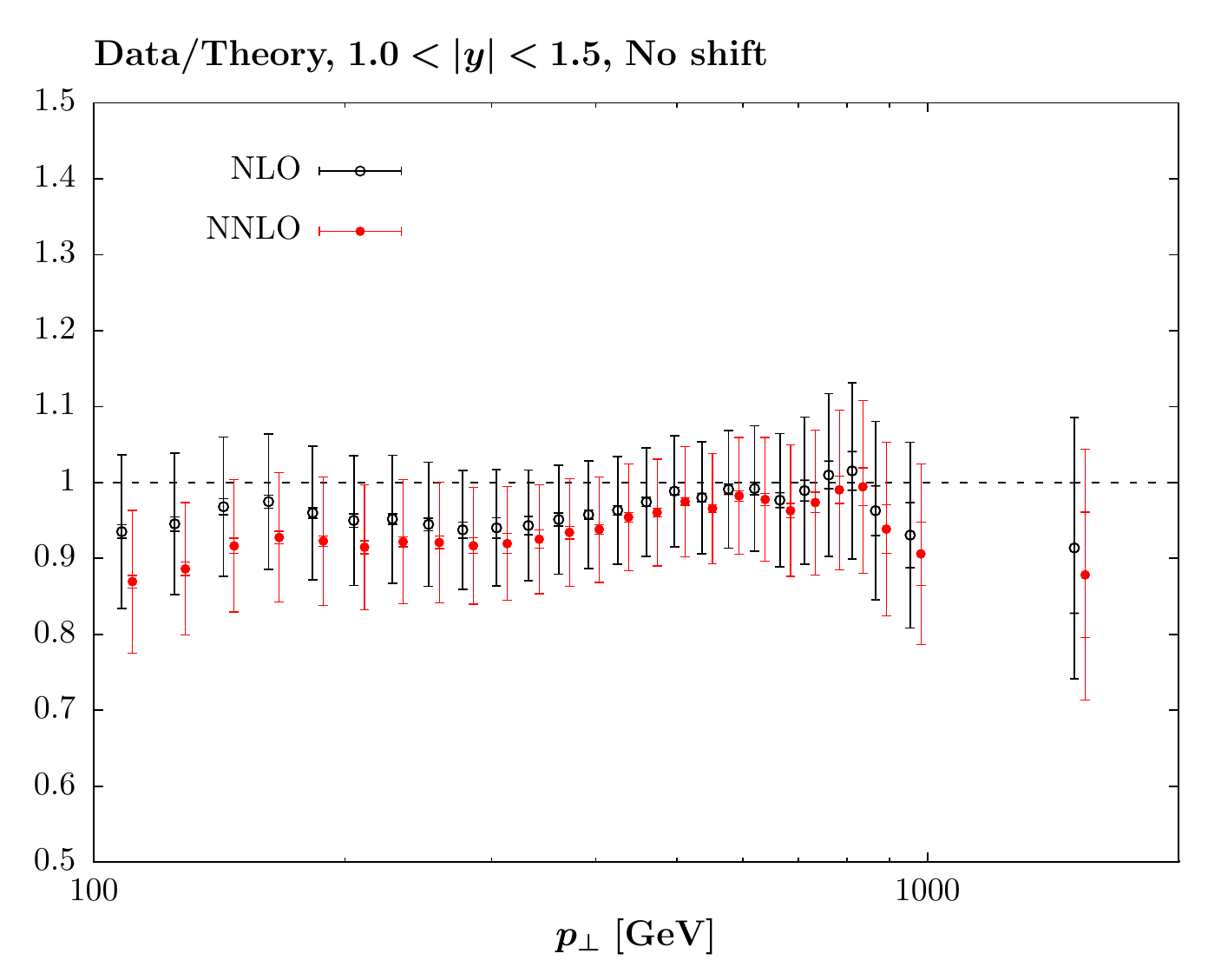}}
\caption{Comparison of NLO prediction and fit to ATLAS jet data~\cite{Aad:2014vwa} for two jet rapidity bins. Data/theory is plotted, without including the shifts due to the systematic uncertainties. Errors are the systematic and statistical added in quadrature.}
\label{fig:jet3}
\end{figure}

\begin{figure}[htb]
\centerline{%
\includegraphics[scale=0.3]{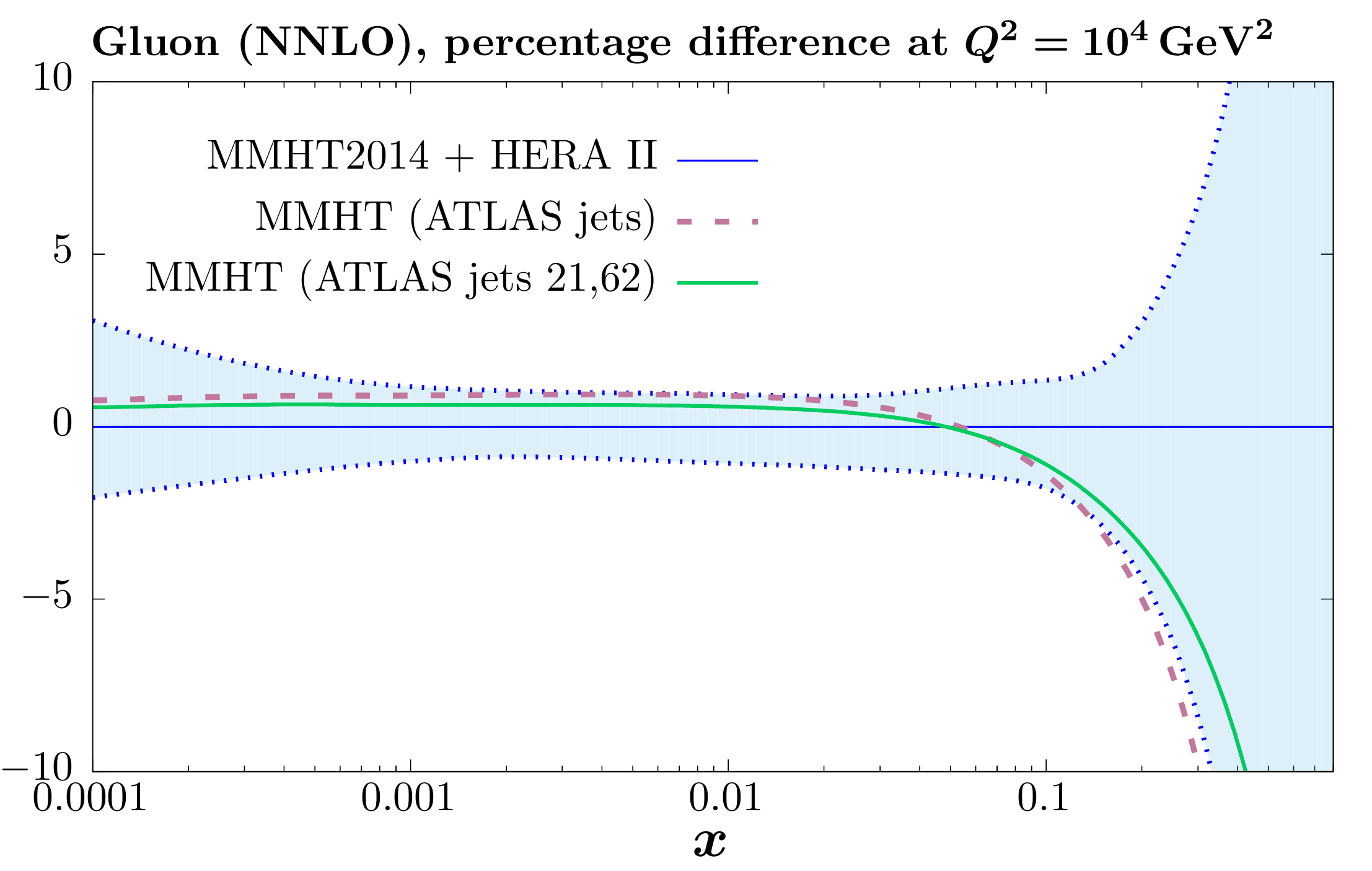}}
\caption{Impact on the gluon PDF of the ATLAS jet data, for the default and systematic error treatment and with errors \texttt{jes21,62}, defined in the text.}
\label{fig:gluon}
\end{figure}

We now consider the impact of including the NNLO corrections calculated in~\cite{Currie:2016bfm} on the data description. The results are shown in Table~\ref{tab:nnlo} for the default treatment of the correlated systematic errors and with the error decorrelation defined in the preceding section. In both cases there is a significant, although not dramatic, deterioration in the $\chi^2$ both with and without the ATLAS data included in the fit.  The source of this effect can be seen most clearly if we consider the data/theory comparison prior to including the shifts due to the correlated systematic errors. This is shown in Fig.~\ref{fig:jet3}, and there is clearly a trend for the NNLO corrections to shift the theory away from the data at lower jet $p_\perp$; such an effect is also visible in the results of~\cite{Currie:2016bfm}. While the final $\chi^2$ will depend on the precise way in which the systematic uncertainties allow the data/theory to shift this will in general lead to some deterioration in the fit quality. This is indeed evident in Table~\ref{tab:nnlo}, and moreover is seen to be roughly independent of the precise treatment of the systematic error correlation. The effect of including the ATLAS data on the gluon PDF is shown in Fig.~\ref{fig:gluon}, and is seen to lead to a somewhat softer gluon at higher $x$, lying on the edge of the PDF uncertainty band. The impact is qualitatively similar, although a little milder, for the decorrelated error treatment. A similar, though smaller, effect is seen at NLO, not displayed here.

In the future, it will be important to consider a wider range of jet radii, $R$, where available. For example, the ATLAS data are presented for a value of $R=0.6$ as well as the case of $R=0.4$ considered here, while the CMS measurement~\cite{Chatrchyan:2012bja} of inclusive jet production at 7 TeV takes $R=0.7$. At the time of these proceedings NNLO K-factors corresponding to these data were not publicly available, however there is some evidence that a large value of $R$ leads to more stable perturbative results~\cite{Currietalk}. In addition, it is worth pointing out that the NLO description of these CMS data, which is included in the MMHT14 fit, is very good, being close to 1 per point~\cite{Harland-Lang:2014zoa}. For these reasons it will therefore be very informative to consider the impact of NNLO corrections on the comparison to the CMS data. A final important factor is the choice of factorization and renormalization scales. For the comparison to the ATLAS data this is given according to the $p_\perp$ of the leading jet in each event. However an alternative choice is to simply treat the data inclusively, taking the $p_\perp$ of each jet in the event as a choice of scale. This is observed to have quite a large impact on the overall result~\cite{Currietalk}. Furthermore, the NLO comparison to the CMS jet data takes such a choice. A complete investigation of all of the above factors will therefore be essential before a full assessment of the impact of NNLO corrections on the comparison to jet data can be made.

\section{Inclusion of new LHC data}

In addition to the final combined HERA I+II data set~\cite{Abramowicz:2015mha}, much new LHC and Tevatron data have become available since the release of the MMHT14 PDF set. We have included a range of these in what we label `MMHT (2016 fit)', an unofficial set that will not be made publicly available but allows the impact of this data on the PDFs to be judged, and paves the way to the public release of a new set. Included in this are the latest $t\overline{t}$ total cross section data, LHCb data~\cite{Aaij:2015gna,Aaij:2015zlq,Aaij:2015vua} on $W$ and $Z$ boson production, CMS data on $W$ boson production~\cite{Khachatryan:2016pev} and $W$ boson production in association with a charm quark~\cite{Chatrchyan:2013uja}, and an updated D0 measurement of the $W\to e\nu$ asymmetry~\cite{D0:2014kma}. In addition, a comparison and fit to the CMS double differential Drell--Yan measurement at 8 TeV~\cite{CMS:2014jea} is attempted, however there are some issues in the comparison that we are currently attempting to resolve. All cross sections are calculated at NLO using MCFM~\cite{Campbell:2015qma} in combination with Applgrid~\cite{Carli:2010rw}, with NNLO K-factors calculated using \texttt{top++}~\cite{Czakon:2011xx} for the $t\overline{t}$ case and FEWZ~\cite{Li:2012wna} for the $W$ and $Z$ case. For $W$ + $c$ production the NNLO calculation is not currently available, so we simply use the NLO calculation in the NNLO fit, as the size of these corrections is expected to be smaller than the experimental uncertainties in the data we compare to.

\begin{table}
\begin{center}
\begin{tabular}{|c|c|c|c|}
\hline
 &Points& NLO $\chi^2$&NNLO $\chi^2$\\ \hline
$\sigma_{t\overline{t}}$ &18&19.6 (20.5)&14.7 (15.3)\\ 
LHCb 7 TeV $W+Z$&33 &50.1 (45.4)&46.5 (42.9)\\
LHCb 8 TeV $W+Z$&34&77.0 (58.9)&62.6 (59.0)\\
LHCb 8 TeV $Z\to ee$ &17&37.4 (33.4)&30.3 (28.9)\\
CMS 8 TeV $W$&22&32.6 (18.6)&34.9 (20.5)\\
CMS 7 TeV $W+c$&10&8.5 (10.0)&8.7 (7.8)\\
D0 $e$ asymmetry&13&22.2 (21.5)&27.3 (25.8)\\
\hline
Total&3405 (3738)&4375.9 (4336.1)&3741.5 (3723.7)\\
\hline
\end{tabular}
\caption{$\chi^2$ at NLO and NNLO for the prediction (fit) to the new LHC and Tevatron data included in the MMHT -- 2016 fit. Also shown is the total number of points without (with) the new data included.}\label{tab:chi}
\end{center}
\end{table}

\begin{figure}[htb]
\centerline{%
\includegraphics[scale=0.7]{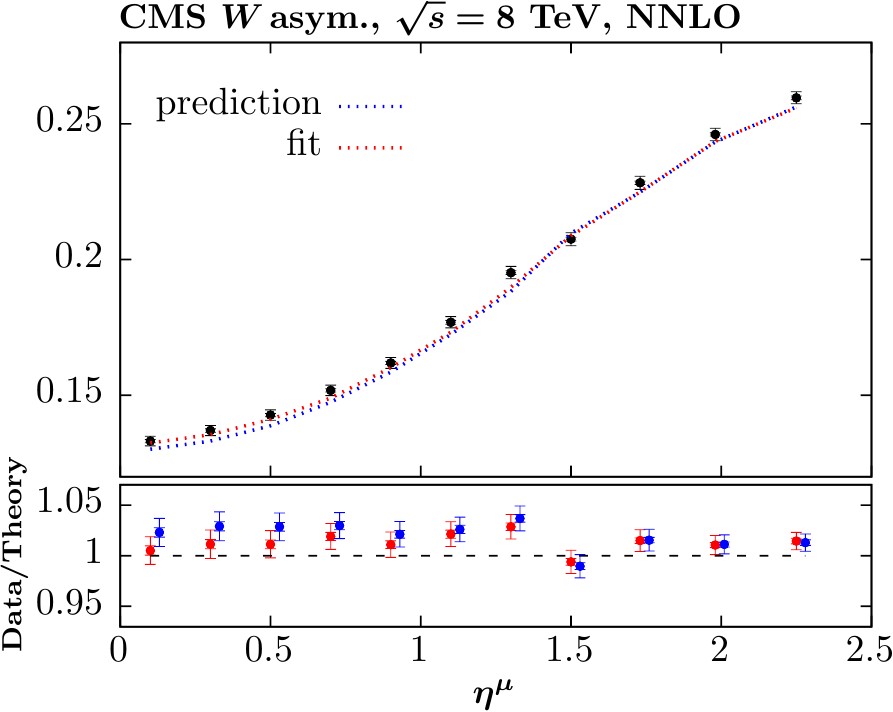}
\includegraphics[scale=0.7]{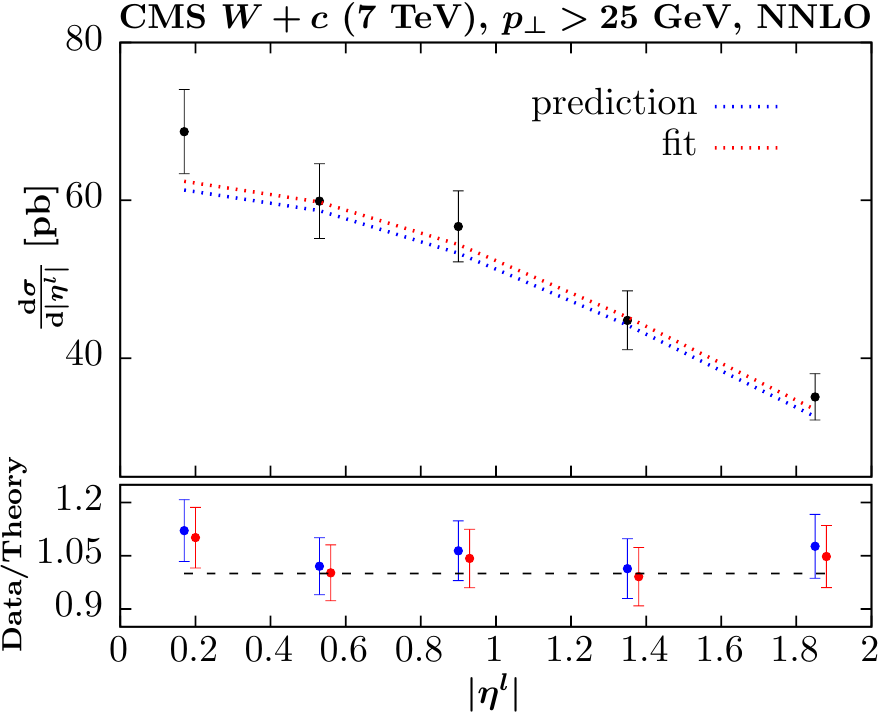}}
\caption{Comparison to CMS $W$ boson~\cite{Khachatryan:2016pev} and $W$ + $c$ production~\cite{Chatrchyan:2013uja} data at NNLO, before and after including the data in the fit. In the former case the $W$ asymmetry is shown for clarity, although the individual $W^\pm$ data are fit to.}
\label{fig:lhcdata}
\end{figure}

The quality of the data description with and without the new data included in the fit at NLO and NNLO is shown in Table~\ref{tab:chi}\footnote{A similar table is shown in~\cite{Harland-Lang:2016zfc}, however there is some small change in the $\chi^2$ values quoted here due to the improved K--factors being used, as well as a more significant change in the case of the CMS $W$ asymmetry data due to a bug, now fixed, in the K--factor implementation.}. The description is generally observed to be good, with some mild improvement after refitting. The one exception to this is the CMS $W$ boson production data~\cite{Khachatryan:2016pev}, where a considerable improvement with refitting is observed. In addition the data description is seen to be somewhat better at NNLO compared to NLO. The best--fit strong coupling $\alpha_S(M_Z^2)$ is found to increase to about 0.118 from 0.1172 at NNLO, while at NLO it remains stable at 0.12. The comparison to the CMS $W$ boson and $W$ + $c$ production data are shown in Fig.~\ref{fig:lhcdata}, and the improvement in the description with refitting for the former case is clear.

\begin{figure}[t]
\centerline{%
\includegraphics[scale=0.5]{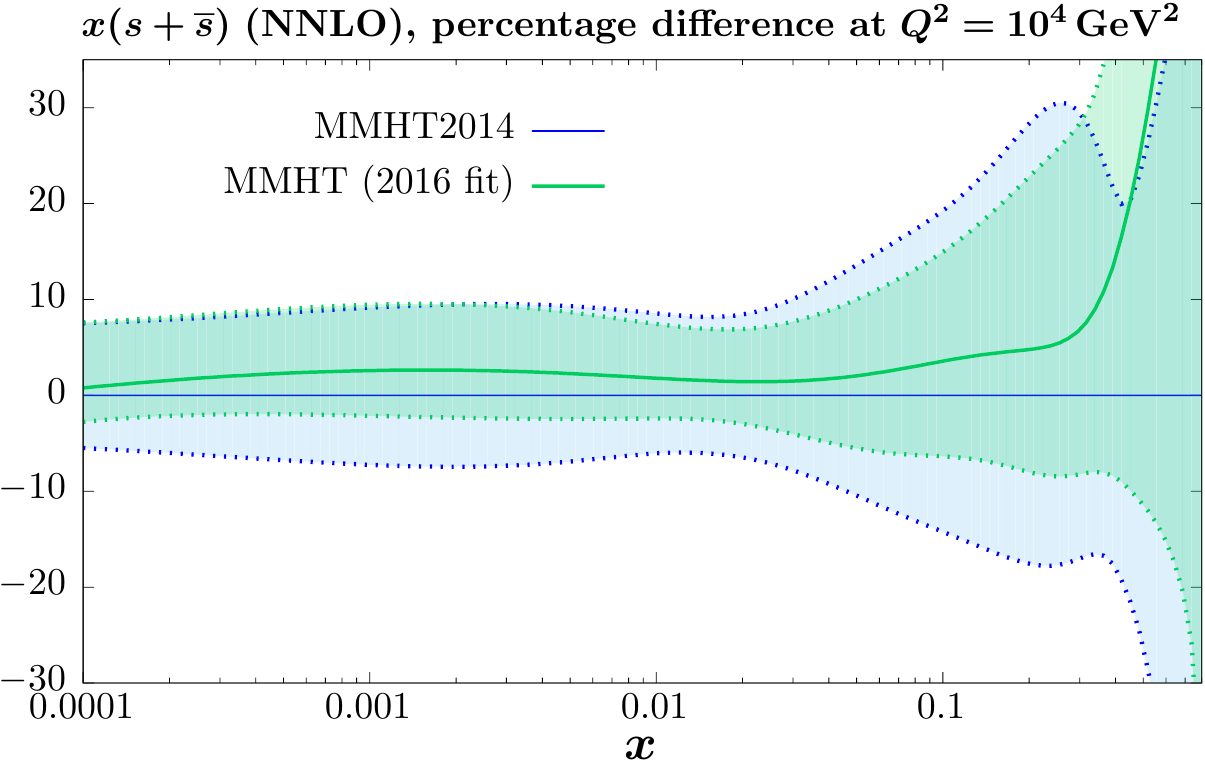}
\includegraphics[scale=0.5]{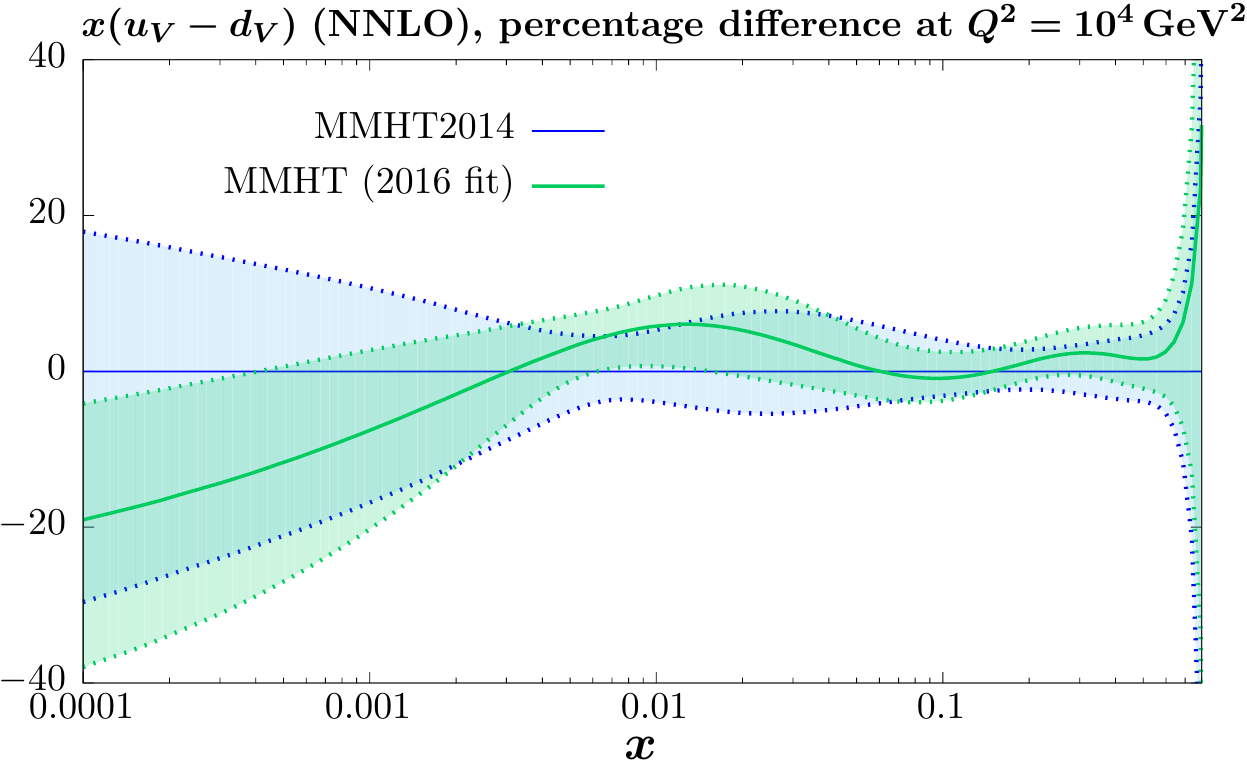}}
\caption{Strange sum $s+\overline{s}$ and valence quark differences $u_V-d_V$ including the 68\% confidence uncertainty bands. Results corresponding to the MMHT14 and MMHT (2016 fit) sets are shown, with the latter case including new LHC data and the final HERA Run--I + II combined data set in the fit.}
\label{fig:lhcpdf}
\end{figure}

As in the case of MMHT14 the `MMHT (2016 fit)' error set has 25 eigenvectors, corresponding to 50 free directions. We find that 14 of these directions are constrained, according to the dynamical tolerance technique described in~\cite{Martin:2009iq}, by the new LHC data. Results for the two most affected PDF combinations, the strange sum $s+\overline{s}$ and valence quark difference $u_V-d_V$ are shown in Fig.~\ref{fig:lhcpdf}. In the former case a significant reduction in the uncertainty is observed, with a mild increase in the central value, due in large part to the CMS $W$ + $c$ data, which is strongly sensitive to this PDF combination. The shape of the valence quark difference changes quite dramatically, with a reduction in the uncertainty at the percent level seen at low and intermediate $x$;  in fact, closer inspection reveals that the dominant change is in fact in the up quark valence distribution. This is mainly driven by the CMS $W$ boson production data, which is sensitive to this quark flavour combination, with some impact coming from the combine HERA data as well. There are some smaller changes in the light sea and gluon PDF, largely driven by the new HERA combined data, which we do not show here for the sake of brevity.

\section{Towards MMHTQED}

While PDFs are more commonly associated with the quarks and gluons within the proton, it is also possible for photon--initiated processes to occur in proton collisions, with a corresponding photon PDF introduced. This is becoming increasingly relevant at the LHC, where NNLO QCD precision is now the standard for a large number of processes. Indeed, as roughly speaking $\alpha(M_Z^2)\sim \alpha_S^2(M_Z^2)$, if we are to quote NNLO QCD accuracy it is crucial to consider the possible contribution from NLO electroweak corrections; photon--initiated processes are one irreducible part of these. Earlier efforts to describe the photon PDF fall into two categories, being either model--dependent attempts based on a simple ansatz due to quark radiation of photons, as in the MRST2004QED~\cite{Martin:2004dh} and more recent CT14QED~\cite{Schmidt:2015zda} sets, or the agnostic treatment of the NNPDFQED set~\cite{Ball:2013hta}, which freely parameterises the photon in the same way as the quark and gluons, with constraints from DIS and LHC $W$ and $Z$ data included. In the latter case this leads to significant uncertainties on the photon, due to the relatively small impact photon--initiated contributions have on such data and hence their limited constraining power. One particular issue that has arisen from the use of this set is the appearance of very large uncertainties in photon--initiated cross sections at high mass, with a central value that can be larger than conventional channels. Such an effect has for example been  discussed in the case of Drell--Yan production in~\cite{Bourilkov:2016qum,Accomando:2016tah,Mangano:2016jyj}, $WW$ production in~\cite{Mangano:2016jyj} and $t\overline{t}$ production in~\cite{Pagani:2016caq}.

More recently, there has been great progress in the determination of the photon PDF, based on the crucial observations that the dominant contribution to the photon is from the well understood elastic $p\to p\gamma$ emission process (see~\cite{Harland-Lang:2016kog} for discussion) and that more generally the photon can be related directly to the proton structure functions probed in $ep$ scattering, which contain both elastic and inelastic contributions, the latter leading into the DIS region as $Q^2$ is increased. This connection is made precise in~\cite{Manohar:2016nzj}, where it is shown that their `LUXqed' photon PDF is generally known with percent level precision in terms of the available structure function data. In particular, they show that the photon can be written as
\begin{align}\nonumber
x\gamma(x,\mu^2)&=\frac{1}{2\pi\alpha(\mu^2)}\int_x^1\frac{{\rm d}z}{z}\Bigg\{\int^{\frac{\mu^2}{1-z}}_{\frac{x^2m_p^2}{1-z}} \frac{{\rm d}Q^2}{Q^2}\alpha^2(Q^2)\\ \nonumber
&\left[\left(zp_{\gamma q}(z)+\frac{2x^2 m_p^2}{Q^2}\right)F_2\left(x/z,Q^2\right)-z^2 F_L\left(x/z,Q^2\right)\right]\\ \label{lux}
&-\alpha^2(\mu^2)z^2 F_2(x/z,\mu^2)\Bigg\}\;,
\end{align}
where $F_{2,L}$ are the usual proton structure functions, and $p_{\gamma q}(z)$ is the LO $\gamma q$ splitting function. While precise, this form relies upon the approximation that the quarks and gluons are independent of the photon, i.e. omitting the impact of the $\gamma \to q\overline{q}$ splitting on the quarks and gluons themselves. While this approximation is generally a good one, with corrections being higher order in $\alpha$, it leads for example to some violation of the momentum sum rule due to the asymmetry in the treatment of the quark/gluons and the photon. In~\cite{Manohar:2016nzj} this is corrected for by absorbing all momentum violation into the gluon PDF, but more generally a full treatment of the coupled DGLAP evolution between the photons and QCD partons, with the input photon PDF at a scale $Q_0$ determined using the same physics input as LUXqed may be preferable. 

\begin{figure}[htb]
\centerline{%
\includegraphics[scale=0.45]{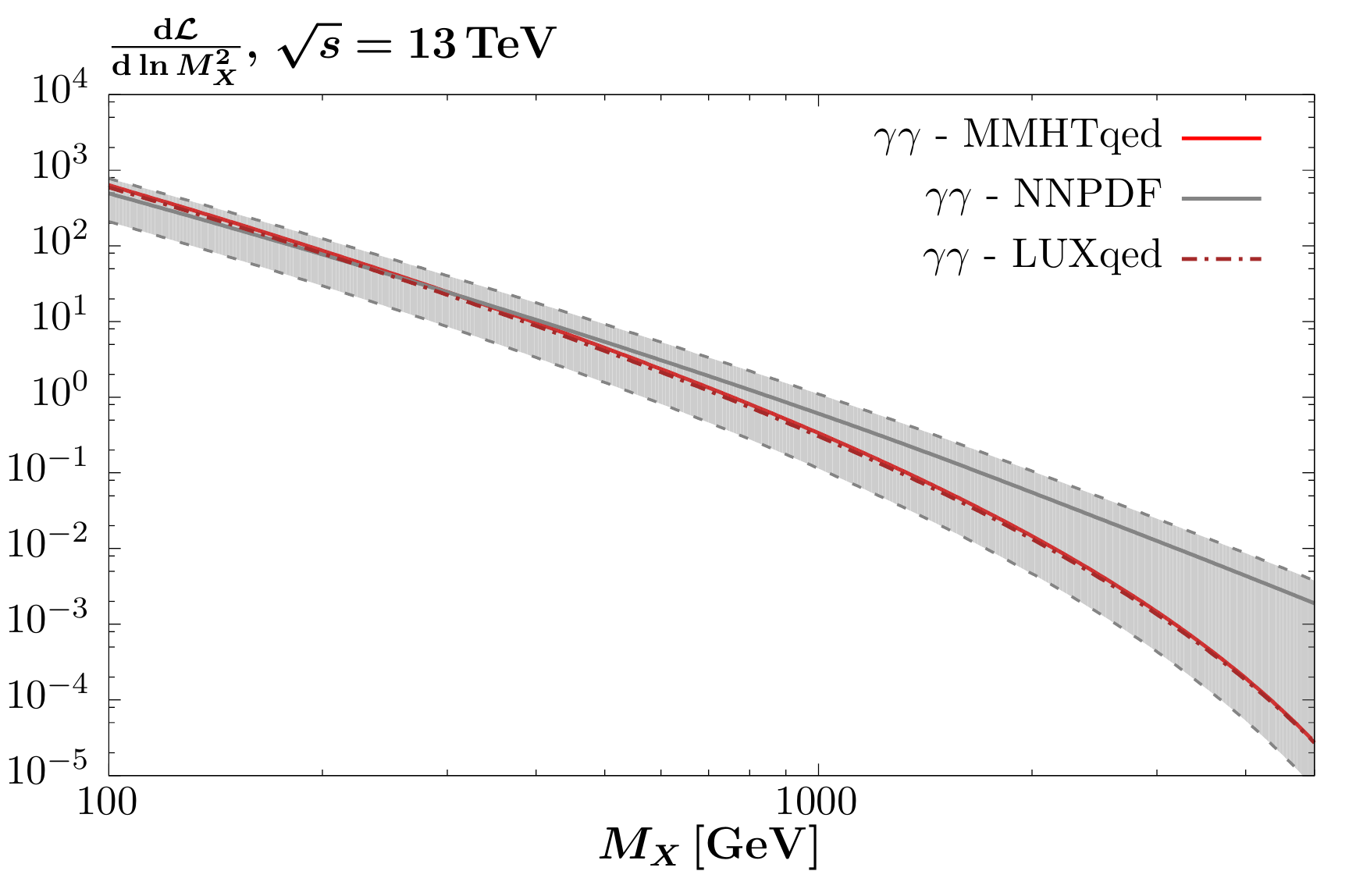}}
\caption{Photon--photon luminosity vs. the invariant mass, $M_X$, of the produced system at 13 TeV, for the NNPDF3.0QED~\cite{Ball:2013hta}, LUXqed~\cite{Manohar:2016nzj} and (preliminary) MMHTQED sets.}
\label{fig:gampdf}
\end{figure}

Work towards including the photon PDF within the MMHT framework is ongoing. In particular, we separate the $Q^2$ integral in (\ref{lux}) into a $Q^2<Q_0^2 = 1$ ${\rm GeV}^2$ region, which determines the input photon $\gamma(x,Q_0^2)$, and a $Q^2>Q_0^2$ region where a suitably modified form of the fully coupled DGLAP evolution is performed within the MMHT framework (work to include the $O(\alpha \alpha_S)$ corrections to the evolution is currently being finalised).  This will allow the photon to be included simply and consistently with future PDF updates. An additional advantage is that the photon PDF of the neutron can be included, with a suitable model of isospin violation at the input scale applied. Finally, it would be possible in principle to include uncertainties on the input due, for example, to the structure functions entering into (\ref{lux}), and allow for the impact of LHC data on for example high mass Drell--Yan production to be  assessed (see~\cite{Giuli:2017oii} for a recent study). The constraining power of such data is unlikely to be competitive, but this will provide a good consistency check.

A first result for the $\gamma\gamma$ luminosity at 13 TeV is shown in Fig.~\ref{fig:gampdf}, with the LUXqed set included for comparison. Broadly speaking very close agreement is seen between the two sets, as expected given the input in the $Q^2 < Q_0^2$ is essentially identical; a more precise comparison is ongoing. Also shown is the NNPDF3.0 prediction, including the corresponding 68\% confidence level uncertainties. These are seen to be very large at higher mass, with the central value being quite high, consistent with the findings of~\cite{Bourilkov:2016qum,Accomando:2016tah,Mangano:2016jyj,Pagani:2016caq}. However, for the updated results for the photon PDF the uncertainties are generally smaller than the line width in the plot, with the central value lying towards to lower end of the NNPDF band. Therefore, we can safely say that there is no room for large photon--initiated contributions with sizeable uncertainties at high mass. Rather, we are now in the era of precision photon PDF phenomenology.

\bibliographystyle{h-physrev}

\bibliography{references}

\end{document}